\documentclass[12pt,letterpaper]{article} 

\usepackage[utf8]{inputenc}
\usepackage[T1]{fontenc}
\usepackage{times} 
\usepackage{amsmath,amsfonts,amssymb}
\usepackage{graphicx}
\usepackage{setspace}
\usepackage[left=0.875in, right=0.875in, top=1.0in, bottom=1.25in]{geometry}

\usepackage{titlesec}
\titleformat{\section}{\normalfont\large\bfseries\uppercase}{\thesection.}{0.5em}{}
\titleformat{\subsection}{\normalfont\large\bfseries}{\thesubsection}{0.5em}{} %

\newcommand{\keywords}[1]{\noindent\textbf{Keywords:} #1}
 
\usepackage{amsmath,amsfonts,amssymb}
\usepackage{graphicx}
\usepackage[colorlinks=true, allcolors=blue]{hyperref}
\usepackage{authblk} 

\usepackage{fancyhdr}
\date{\today}        
\date{Octobre 2026}  

\title{From Early Concepts to Lunar Hypertelescopes Roadmap for Next-Generation Optical Astronomy on the Moon}

\author[a,b,c]{M. Hadjara}
\author[a]{T. Xu}
\author[d]{S. Lagarde}
\author[a]{H. Ji}
\author[a]{Q. Dong}
\author[c]{M. Diaz}
\author[e]{M. Meftah}
\author[f,g]{B. Foing}
\author[h]{Z. Ping}
\author[i]{X. Chen}
\author[j]{R. Mendez}
\affil[a]{Nanjing Institute of Astronomical Optics \& Technology, National Astronomical Observatories, Chinese Academy of Sciences, Nanjing 210042, China}
\affil[b]{Chinese High Angular Resolution Southern Astronomical Laboratory (CHARSAL), Dept. of Astronomy,  Camino El Observatorio \#1515, Las Condes, Santiago, Chile. Postal Code: 7591245. Casilla 36-D, Santiago - Chile. (E-mail: mhadjara@uchile.cl)}
\affil[c]{Space and Planetary, Exploration Laboratory (SPEL), Dept. of Electrical Engineering, FCFM, Universidad de Chile, Av. Tupper 2007, Santiago, Chile}
\affil[d]{Université de la Côte d’Azur, Observatoire de la Côte d’Azur, CNRS, Laboratoire Lagrange, Bd de l’Observatoire, CS 34229, 06304 Nice cedex 4, France}
\affil[e]{LATMOS (Laboratoire Atmosphères, Milieux, Observations Spatiales), Université Paris-Saclay, France}
\affil[f]{ILEWG (International Lunar Exploration Working Group), Vrije Universiteit Amsterdam, The Netherlands}
\affil[g]{ERA chair for space photonics , Latvia U , Riga and Leiden Observatory}
\affil[h]{Institute for Advanced Study, Shenzhen University, Shenzhen, China}
\affil[i]{Shanghai Astronomical Observatory, Chinese Academy of Sciences, Shanghai, China}
\affil[j]{Departamento de Astronomía (DAS), Universidad de Chile, Santiago, Chile}

\begin{document} 
\maketitle

\thispagestyle{fancy} 
\pagestyle{fancy}     

\begin{abstract}
Since the dawn of the Space Age, the Moon has been considered one of the most promising locations for astronomical observations. The absence of atmosphere, exceptional mechanical stability, access to the permanently shielded radio-quiet far side, and the possibility of constructing extremely large apertures have motivated numerous concepts for lunar observatories over the past six decades. Despite extensive studies conducted during the Apollo era, most proposed lunar astronomy projects remained unrealized due to technological limitations, shifting scientific priorities, and the high cost of sustained lunar operations.
Today, renewed international interest in lunar exploration, combined with advances in robotics, autonomous systems, additive manufacturing, precision optics, and formation flying, is reopening the prospect of large-scale astronomical infrastructures on the Moon. In particular, lunar hypertelescope concepts offer the possibility of achieving unprecedented angular resolutions through diluted optical arrays spanning hundreds of meters to several kilometres.
This paper reviews the historical evolution of lunar astronomy concepts, analyses the reasons behind previous failures, examines current technological developments, and proposes a roadmap toward future lunar hypertelescope facilities capable of transforming observational astronomy during the twenty-first century.
\end{abstract}


\section{INTRODUCTION}
\label{sec:intro}  

The Moon has occupied a unique place in human imagination since antiquity. Long before becoming a destination for scientific exploration, it inspired mythology, religion, philosophy, literature, and early astronomy.
The invention of the telescope transformed humanity's perception of the Moon. Galileo's telescopic observations of the Moon in 1609 revealed mountains, valleys, and other irregular surface features, challenging the Aristotelian view of celestial perfection. These discoveries initiated a profound transformation in natural philosophy that culminated with Newton's law of universal gravitation, demonstrating that the same physical laws govern both terrestrial and celestial phenomena \cite{galilei1610sidereus,north2008cosmos,newton1687principia}. Over subsequent centuries, lunar observations became central to the development of modern astronomy.

During the Industrial Revolution and the rapid development of modern engineering, humanity began to dream of travelling to the Moon, a vision reflected in literature (e.g., From the Earth to the Moon by Jules Verne, 1865) and early cinema (e.g., A Trip to the Moon by Georges Méliès, 1902). This dream became reality on July 20, 1969, with the successful landing of the Apollo 11 mission on the lunar surface. Since then, the Space Age has transformed the Moon from a distant object of observation into a potential platform for scientific exploration. Following the Apollo program, scientists began to consider the establishment of astronomical observatories on the lunar surface \cite{compton1989apollo}.

Although several preliminary studies have remained conceptual, they laid the foundation for a number of ideas that are now being actively revisited in the context of renewed lunar exploration programs and advances in space instrumentation. In particular, the Moon is increasingly reconsidered not only as a scientific target, but also as a potential platform for future astronomical observatories, motivated by its unique physical and environmental properties discussed in the following section \cite{lowman1966lunarresources,Foing1994,Foing1994Proc}.

\section{WHY THE MOON IS AN EXCEPTIONAL SITE FOR ASTRONOMY}
\label{sec:2}

The Moon offers a combination of environmental and mechanical conditions that make it an outstanding platform for future astronomical facilities.

A first key advantage is the absence of a substantial atmosphere. On Earth, observational performance is fundamentally limited by:
\begin{itemize}
\item atmospheric turbulence (seeing degradation),
\item absorption and scattering across most wavelength ranges,
\item airglow and sky background emission.
\end{itemize}
These effects require complex mitigation strategies such as adaptive optics, site selection at high altitude, or space deployment. In contrast, the lunar surface provides an atmosphere-free environment, enabling intrinsically diffraction-limited observations without atmospheric perturbations or sky background contamination.

A second advantage is the mechanical stability of the lunar surface. In the absence of environmental disturbances such as:
\begin{itemize}
\item wind loading,
\item weather effects,
\item atmospheric pressure fluctuations,
\item low-level seismic activity (moonquakes),
\end{itemize}
the Moon provides a rigid and stable platform for precision optical instrumentation. This stability is particularly relevant for long-baseline interferometry and hypertelescope architectures, where nanometric path-length stability and long-term structural coherence are required.

A third unique property is the radio-quiet environment, especially on the lunar far side. This region is naturally shielded from:
\begin{itemize}
\item terrestrial radio-frequency interference,
\item ionospheric plasma effects that limit low-frequency observations from Earth.
\end{itemize}
As a result, it provides access to an exceptionally clean electromagnetic environment, enabling ultra-low-frequency radio astronomy that is otherwise inaccessible from ground-based observatories.

Finally, the lunar surface enables the deployment of extremely large and mechanically supported apertures. Unlike free-flying space telescopes constrained by launch vehicle fairing sizes and formation-flying complexity, lunar-based systems may exploit the terrain directly to support large structures. This opens the possibility of interferometric arrays with baselines ranging from hundreds of meters to kilometers, enabling angular resolutions far beyond current ground- and space-based facilities.

\section{Historical Lunar Astronomy Concepts (1960--1970s)}
\label{sec:3}

The Apollo era triggered a strong interest in deploying astronomical instruments on the Moon, leading to a series of conceptual and preliminary mission studies. These early ideas explored the Moon as a stable platform for optical, ultraviolet, infrared, and radio astronomy, benefiting from the absence of atmosphere and, in some cases, radio shielding on the lunar far side.

Table~\ref{Tab:1} summarizes the main historical concepts of lunar-based astronomical facilities proposed during the 1960s and 1970s.

\begin{table}[h!]
\centering
\renewcommand{\arraystretch}{1.2}
\begin{tabular}{p{3.5cm} p{1.5cm} p{2.5cm} p{4.3cm} p{3.3cm}}
\hline
\textbf{Concept / Project} & \textbf{Country} & \textbf{Period} & \textbf{Objective} & \textbf{Status} \\
\hline

Lunar Optical Telescope & USA & $\sim$1966--1970 &
High-resolution optical imaging from the lunar surface &
Conceptual study \\

Lunar Surface UV Telescope (LSUT) & USA & $\sim$1968--1972 &
Ultraviolet observations free from atmospheric absorption &
Study linked to Apollo planning \\

\textbf{Apollo Far-Ultraviolet Camera/Spectrograph (UVC)} & \textbf{USA} & \textbf{1972} \textbf{(Apollo 16)} & \textbf{UV imaging of stars and Earth’s upper atmosphere} & \textbf{First astronomical instrument operated on the Moon} \\

Lunar-Based Astronomical Observatories & USA & $\sim$1968--1975 &
Permanent or semi-permanent observatories (optical/UV/IR) on the lunar surface &
Conceptual only \\

ALSEP Extensions (astronomy payloads) & USA & 1971--1972 &
Possible integration of infrared or X-ray sensors into ALSEP infrastructure &
Not implemented \\

Lunokhod UV/X-ray Payload Concepts & USSR & $\sim$1970--1973 &
Addition of astronomical sensors to robotic lunar rovers &
Conceptual \\

Automated Lunar Infrared Observatories & USSR & $\sim$1970 &
Early studies of remotely operated IR telescopes on the lunar surface &
Conceptual \\

Cryogenic Lunar Infrared Telescope (polar sites) & USA & 1970s &
Use of permanently shadowed regions for deep IR/sub-mm observations &
Theoretical proposal \\

\textbf{Lunar Far-Side Radio Astronomy Arrays} & USA / Global & 1960s onward &
Radio observations shielded from terrestrial interference &
Proposed concept, later \textbf{revived} (e.g., Artemis era) \\

\hline
\end{tabular}
\caption{Historical concepts of lunar-based astronomical observatories proposed during the Apollo era. Only the Apollo 16 Far-Ultraviolet Camera/Spectrograph was successfully implemented, becoming the first astronomical instrument operated on the Moon.}
\label{Tab:1}
\end{table}

Among these early concepts, only the Apollo Far-Ultraviolet Camera/Spectrograph (UVC) was successfully deployed. Operated during Apollo 16, it provided the first astronomical observations from the lunar surface, imaging both stellar sources and Earth’s upper atmosphere in the far-ultraviolet domain.

\section{Why Lunar Astronomy Did Not Materialize}
\label{sec:4}

Despite the strong conceptual interest generated during the Apollo era, lunar-based astronomy did not progress beyond early studies. Several converging factors explain this outcome:

\begin{itemize}
\item \textbf{Shift toward Earth-orbit observatories:}  
After the Apollo program, space agencies progressively prioritized space-based telescopes in Earth orbit. Missions such as the \textit{Hubble Space Telescope}, \textit{Chandra X-ray Observatory}, and \textit{Spitzer Space Telescope} demonstrated that high-impact astronomy could be achieved without the operational complexity of lunar deployment. These facilities offered immediate scientific return, servicing capability (in some cases), and significantly reduced mission risk compared to lunar surface installations.

\item \textbf{Technological limitations:}  
At the time, several key enabling technologies required for sustainable lunar observatories were not yet mature. These included autonomous robotic systems capable of long-duration scientific operations, high-precision deployment and alignment mechanisms, lightweight and stable optical structures, reliable long-term power generation and storage, and effective strategies for mitigating lunar dust contamination. Together, these limitations made complex astronomical installations on the lunar surface technically premature.

\item \textbf{Economic and political constraints:}  
Following the Apollo era, political momentum and financial commitment toward large-scale lunar infrastructure declined significantly. Priority shifted toward cost-effective, incremental missions in low-Earth orbit. As a result, the scientific community increasingly favored orbital observatories, which provided a more achievable balance between cost, risk, and scientific return.
\end{itemize}

\section{THE RETURN OF LUNAR ASTRONOMY}
\label{sec:5}

The situation has changed dramatically during the last decades. Several developments now favour a renewed interest in lunar observatories.

\subsection{International Lunar Programs}
\label{sec:5.1}

Unlike the situation that prevailed during the Apollo era, several large-scale lunar exploration programs are currently under development and are expected to provide the infrastructure required for future astronomical facilities on the Moon.

NASA's Artemis program aims to establish a sustained human presence on and around the Moon through a series of crewed and robotic missions, including the Lunar Gateway orbital platform and long-duration surface operations \cite{NASAArtemis2023}.

In parallel, China and Russia have proposed the International Lunar Research Station (ILRS), a long-term scientific infrastructure envisioned for robotic and eventually human operations at the lunar surface \cite{Wu2021ILRS, CNSA2021}.

Commercial participation is also rapidly expanding through NASA's Commercial Lunar Payload Services (CLPS) initiative, which supports the delivery of scientific instruments and technology demonstrators to the lunar surface by private companies \cite{CLPS2024}.

These programs collectively represent the emergence of a permanent cislunar ecosystem that could dramatically reduce the cost and complexity of deploying astronomical observatories on the Moon.

\subsection{Recent Astronomical Demonstrators and Lunar Observatory Concepts}
\label{sec:5.2}

Beyond the development of lunar exploration infrastructure, several dedicated astronomical projects have emerged during the last decade, demonstrating a renewed scientific interest in lunar-based observations.

One of the most significant achievements is the Lunar-based Ultraviolet Telescope (LUT), deployed aboard China's Chang'e-3 mission in 2013. Operating in the near-ultraviolet wavelength range between 245 and 340 nm, LUT became the first operational astronomical telescope on the lunar surface. Since its deployment, the instrument has successfully monitored numerous variable stars and transient sources, demonstrating the long-term stability and scientific viability of lunar astronomical observations \cite{Wang2015LUT,Cao2017LUT}.

Radio astronomy has also experienced a resurgence of interest through several ambitious lunar concepts. Among them, the Lunar Crater Radio Telescope (LCRT), studied under NASA's Innovative Advanced Concepts (NIAC) program, proposes the construction of a giant radio telescope approximately 1.3 km in diameter inside a naturally occurring lunar crater on the far side of the Moon \cite{Burns2021LCRT}. By exploiting the crater as a structural support, the concept minimizes construction requirements while taking advantage of the exceptionally radio-quiet environment provided by the lunar far side.

Complementing surface-based radio observatories, the Dark Ages Polarimeter Pathfinder (DAPPER) mission concept aims to place a radio observatory in lunar orbit. Operating in the frequency range of approximately 40--120 MHz, DAPPER is designed to investigate the cosmic Dark Ages and Cosmic Dawn through measurements of the redshifted 21 cm hydrogen signal \cite{Burns2020DAPPER}. Such observations are extremely challenging from Earth due to ionospheric distortions and radio-frequency interference.

The lunar poles have attracted particular attention for infrared astronomy. Permanently shadowed craters may maintain temperatures below 40 K, creating natural cryogenic environments that could substantially reduce thermal background noise for infrared observations \cite{Lester2020,Angel2008}. Several studies conducted by NASA, ESA, and the broader astronomical community have explored the possibility of deploying infrared observatories in these regions to investigate primordial galaxies, star formation processes, and exoplanet atmospheres.

These projects span a wide range of wavelengths, from radio to ultraviolet and infrared astronomy, illustrating the growing maturity of lunar astronomy concepts. More importantly, they demonstrate that lunar observatories are no longer limited to theoretical studies but are progressively evolving toward operational scientific facilities.

\subsection{Robotics and Automation}
\label{sec:5.3}

Robotic technologies have advanced considerably since the early lunar astronomy studies of the 1960s and 1970s.

Modern autonomous systems are capable of navigation, site preparation, instrument deployment, maintenance operations, and scientific observations with minimal human intervention \cite{Fong2008, Biesiadecki2020}.

Artificial intelligence, machine vision, autonomous planning, and teleoperation technologies developed for planetary exploration missions now offer the possibility of assembling and operating complex observatories in harsh lunar environments. Such capabilities are particularly attractive for facilities located in remote regions, including permanently shadowed craters or the lunar far side.

\subsection{Additive Manufacturing}
\label{sec:5.4}

One of the most transformative developments for future lunar infrastructure is the growing maturity of in-situ resource utilization (ISRU) and additive manufacturing technologies.

Rather than transporting all structural materials from Earth, future lunar facilities may exploit local regolith to manufacture landing pads, roads, protective shielding, and even telescope support structures \cite{Anand2012, SandersLarson2015}.

Several studies have demonstrated the feasibility of producing construction elements from lunar regolith simulants using sintering, microwave processing, and three-dimensional printing techniques \cite{Cesaretti2014, Meurisse2018}.

These approaches could substantially reduce launch mass requirements while enabling the construction of larger and more ambitious astronomical facilities than would otherwise be economically feasible.

\subsection{Advanced Optical Technologies}
\label{sec:5.5}

Optical engineering has undergone remarkable progress during the past several decades, significantly expanding the range of astronomical facilities that could be deployed on the lunar surface.

Recent developments include lightweight segmented mirrors, active and adaptive optics systems, integrated photonic instrumentation, precision laser metrology, and advanced optical interferometry techniques \cite{Angel2003, Lawson2000, BlandHawthorn2019}.

Large deployable mirrors have already been demonstrated in space observatories such as the James Webb Space Telescope, while precision formation flying and interferometric concepts continue to mature for future missions \cite{Gardner2006, Mather2022}.

Integrated photonic technologies are also revolutionizing astronomical instrumentation by reducing mass, volume, and alignment complexity while increasing stability and performance \cite{BlandHawthorn2019}.

Together, these advances significantly reduce many of the technological barriers that limited earlier lunar telescope concepts and open new possibilities for large-scale optical and infrared observatories on the lunar surface.

\subsection{New Generation of Advanced Power Supply}
\label{sec:5.6}

Recent developments in advanced nuclear energy systems suggest new pathways for long-duration, high-autonomy power generation in extraterrestrial environments. Among these, molten salt reactors (MSRs) based on thorium fuel cycles are gaining renewed interest due to their inherent safety features, high fuel utilization efficiency, and reduced long-lived radioactive waste production.

In particular, the Chinese Thorium Molten Salt Reactor (TMSR-LF1), developed by the Shanghai Institute of Applied Physics (SINAP), has recently achieved a key milestone with the first experimental thorium-to-uranium fuel conversion in an operational MSR configuration \cite{WNN_thorium_2025,CAS_thorium_2025}. The 2 MWt experimental reactor, located in Wuwei (Gansu Province), reached first criticality in 2023 and has since entered an operational testing phase under regulatory approval \cite{WNN_operational_2023}.

The long-term roadmap of this program includes the development of higher-power demonstrators (up to $\sim$100 MWt by 2035), with potential scalability toward industrial systems capable of supporting remote infrastructures, including lunar surface installations \cite{WNN_thorium_2025}. In parallel, strategic studies have discussed the use of compact nuclear systems for sustained lunar operations, including integration within International Lunar Research Station (ILRS) scenarios targeting the 2035--2050 horizon \cite{Moon_Nuclear_ILRS_2025}.

Such advanced nuclear power systems provide a credible enabling technology for future lunar observatories, particularly for energy-intensive infrastructures such as interferometric arrays, cryogenic systems, and long-delay-line optical architectures operating in permanently shadowed or high-latitude regions of the Moon.

\subsection{Why the 2020s May Finally Enable Lunar Astronomy}
\label{sec:5.7}

The history of lunar astronomy reveals a recurring paradox. For more than half a century, astronomers have recognized the exceptional scientific potential of the Moon as a platform for astronomical observations. Yet despite numerous studies and conceptual designs, no major lunar observatory has ever been realized.

The primary reason was not a lack of scientific motivation, but rather the absence of a sustainable lunar infrastructure. During the Apollo era, access to the lunar surface remained extremely expensive, infrequent, and dependent on short-duration human missions. Following Apollo, scientific priorities shifted toward Earth-orbiting observatories, which offered lower costs, greater flexibility, and faster scientific returns. Today, however, the situation is fundamentally different.

Multiple international programs, including Artemis, the International Lunar Research Station (ILRS), and commercial lunar initiatives, are establishing the foundations of a long-term cislunar economy and a permanent lunar presence. For the first time since Apollo, regular access to the lunar surface appears technically and politically realistic \cite{NASAArtemis2023,Wu2021ILRS,CLPS2024}.

At the same time, major advances in robotics and autonomous systems are reducing the need for continuous human intervention. Modern robotic platforms can perform deployment, assembly, maintenance, and scientific operations with a level of autonomy unimaginable during the early lunar studies of the 1960s and 1970s \cite{Fong2008,Biesiadecki2020}.

The emergence of in-situ resource utilization (ISRU) and additive manufacturing technologies further changes the equation. Future observatories may be able to exploit local resources for construction, shielding, and infrastructure development, significantly reducing launch mass and overall mission costs \cite{Anand2012,Cesaretti2014,Meurisse2018}.

Perhaps equally important are the extraordinary advances achieved in astronomical instrumentation itself. Lightweight segmented mirrors, active optics, astrophotonics, precision metrology, and optical interferometry have dramatically expanded the range of observatory architectures that can now be envisioned for lunar deployment \cite{Angel2003,BlandHawthorn2019,Lawson2000}.

Recent progress in advanced space power systems, particularly high-efficiency nuclear technologies such as compact fission reactors and molten-salt thorium-based systems, also contributes to this paradigm shift. These systems are increasingly considered as viable candidates for providing continuous, high-power energy supply for sustained lunar operations, including demanding infrastructures such as cryogenic facilities, long-baseline interferometers, and autonomous robotic networks \cite{WNN_operational_2023, Moon_Nuclear_ILRS_2025}.

Taken individually, none of these developments would necessarily guarantee the success of lunar astronomy. Taken together, however, they represent a profound change in the technological landscape. The obstacles that prevented lunar astronomy from materializing during the twentieth century are progressively being removed.

As a result, the question facing the astronomical community may no longer be whether lunar observatories are feasible, but rather which scientific objectives uniquely justify their deployment and how they can be integrated into the broader ecosystem of ground-based and space-based observatories.

The coming decades may therefore represent the first realistic opportunity to transform lunar astronomy from a recurring vision into an operational scientific reality.

\subsection{Key Challenges for Lunar Astronomy}
\label{sec:5.8}

Despite the remarkable scientific opportunities offered by the lunar environment, several major technological and operational challenges must still be addressed before large-scale lunar observatories can become a reality.

\subsubsection{Logistics and Maintenance}
\label{sec:5.8.1}

The transportation of large and fragile astronomical instruments to the lunar surface remains costly, complex, and inherently risky. Unlike ground-based observatories, lunar facilities cannot currently rely on routine human intervention for maintenance, repairs, upgrades, or troubleshooting. Although robotic systems are becoming increasingly capable, autonomous deployment and assembly of large optical infrastructures remain challenging due to the precision required for astronomical instrumentation.

\subsubsection{Lunar Dust (Regolith)}
\label{sec:5.8.2}

Lunar regolith represents one of the most significant environmental hazards for long-term astronomical operations. The dust particles are extremely fine, abrasive, and often electrostatically charged. As demonstrated during the Apollo missions, lunar dust readily adheres to exposed surfaces, including optics, mechanical joints, radiators, solar panels, and electronic components. For astronomical observatories, mirror contamination and degradation of optical surfaces could significantly reduce instrument performance, increase maintenance requirements, and shorten operational lifetimes.

\subsubsection{Power Supply and Energy Management}
\label{sec:5.8.3}

Reliable energy production remains a critical requirement for any lunar observatory. Most locations on the lunar surface experience approximately fourteen Earth days of continuous darkness during each lunar night, severely limiting the effectiveness of solar power systems. Future observatories will likely require a combination of solar arrays, advanced battery storage, fuel cells, or compact nuclear power systems to ensure uninterrupted operations. Continuous electrical power is particularly important for cryogenic instruments, active optics, communication systems, data transmission, and autonomous robotic infrastructure.

\subsubsection{Pointing, Tracking, and Orientation}
\label{sec:5.8.4}

Although the absence of an atmosphere eliminates wind-induced disturbances and atmospheric turbulence, precise telescope pointing and tracking remain essential. Scientific observations often require sub-arcsecond or even milli-arcsecond stability over extended periods. Consequently, lunar observatories must incorporate highly accurate motorized pointing systems, active stabilization mechanisms, and autonomous control algorithms. Furthermore, local illumination conditions, terrain constraints, and the slow lunar rotation introduce additional operational complexities that must be considered during system design.

\subsubsection{Thermal Control and Cooling}
\label{sec:5.8.5}

Thermal management is among the most demanding engineering challenges on the Moon. Surface temperatures can vary from approximately $-170^\circ$C during lunar night to more than $+120^\circ$C under direct sunlight. Such extreme thermal cycles can induce structural deformations, optical misalignments, and degradation of sensitive electronics. Many future astronomical instruments, particularly infrared detectors, will require cryogenic operating temperatures. Potential solutions include passive radiative cooling, deployment within permanently shadowed regions near the lunar poles, underground installation, and the use of dedicated thermal shielding and radiator systems.

Despite these challenges, ongoing advances in robotics, autonomous systems, energy generation, precision engineering, and lunar infrastructure development suggest that many of these obstacles may be progressively mitigated, as outlined in previous subsections, during the coming decades.

\section{PASSIVE LUNAR TELESCOPES USING NATURAL CRATERS}
\label{sec:6}

One particularly attractive approach for future lunar astronomy consists of exploiting natural lunar topography as an integral component of the observatory architecture.

Unlike terrestrial observatories, which generally require extensive civil engineering and structural support systems, the lunar surface offers a vast number of impact craters that may serve as pre-existing foundations for large astronomical instruments. By taking advantage of naturally occurring depressions, it may be possible to construct observatories with apertures far larger than those achievable through conventional engineering approaches.

Large lunar craters are particularly attractive for radio astronomy. Their geometry can provide natural support for suspended receiver systems while reducing the amount of structural material that must be transported from Earth. This concept has been extensively investigated in the context of the Lunar Crater Radio Telescope (LCRT), a proposed kilometer-scale radio observatory to be deployed within a crater on the lunar far side \cite{Burns2021LCRT}.

The lunar far side is especially advantageous for low-frequency radio astronomy because it is naturally shielded from terrestrial radio-frequency interference and ionospheric distortions. As a result, it represents one of the quietest electromagnetic environments accessible within the inner Solar System \cite{Burns2013FARSIDE, Jester2009}.

Similar concepts have also been proposed for optical and infrared astronomy. Lunar craters may serve as stable platforms for supporting large reflective surfaces or segmented mirror assemblies. In some cases, the crater geometry itself can contribute to the overall optical design, reducing the complexity of the supporting infrastructure.

An influential example is the concept of lunar liquid mirror telescopes proposed by Angel and collaborators \cite{Angel2008}. These studies demonstrated that extremely large zenith-pointing telescopes, potentially reaching apertures of tens to hundreds of meters, could be constructed on the Moon using rotating reflective liquids operating under lunar gravity \cite{Borra2007}.

The use of natural terrain offers several potential advantages:

\begin{itemize}
\item Reduced construction and transportation requirements;
\item Access to extremely large apertures;
\item Long-term mechanical stability due to the absence of atmospheric weathering;
\item Efficient utilization of existing geological structures;
\item Potential reduction of deployment costs through robotic installation.
\end{itemize}

Recent advances in autonomous robotics, lightweight materials, precision deployment systems, and in-situ resource utilization have renewed interest in these concepts. Several studies have proposed transforming suitably shaped craters into giant reflective structures through robotic deployment of metallic meshes, thin-film reflectors, or tensioned membrane systems \cite{Burns2021LCRT, Zacny2019}.

Although significant technological challenges remain, passive crater-based observatories represent one of the most promising pathways toward achieving kilometer-scale collecting areas on the lunar surface. Such facilities could enable scientific observations that are currently beyond the reach of both terrestrial and space-based observatories.

\section{FROM TELESCOPES TO HYPERTELESCOPES}
\label{sec:7}

While large monolithic telescopes remain attractive for many scientific applications, the most ambitious long-term vision for lunar astronomy may be the deployment of giant optical interferometric arrays, commonly referred to as hypertelescopes.

Originally proposed by Labeyrie, the hypertelescope concept seeks to overcome the practical limitations associated with constructing increasingly larger monolithic mirrors. Instead of relying on a single collecting surface, the instrument combines light gathered by a large number of individual apertures distributed over vast distances \cite{Labeyrie1996}.

A key innovation of the hypertelescope is the use of pupil densification, a beam-combination technique that preserves the angular resolution associated with the maximum baseline while concentrating the light into a directly usable image. This approach differs from conventional stellar interferometry, which often relies on indirect image reconstruction from visibility measurements \cite{Labeyrie1996, Labeyrie2013}.

Unlike traditional monolithic telescopes, hypertelescopes rely on:

\begin{itemize}
\item Diluted apertures distributed over large baselines;
\item Coherent optical beam combination;
\item Densified pupil imaging;
\item Precision optical path control and metrology.
\end{itemize}

The achievable angular resolution scales approximately as:

\begin{equation}
\theta \approx \frac{\lambda}{B},
\end{equation}

where $\lambda$ is the observing wavelength and $B$ is the maximum baseline of the array.

Consequently, the angular resolution is determined primarily by the separation between the outermost apertures rather than by the size of the individual collecting elements. For baselines approaching several kilometers, angular resolutions may reach microarcsecond levels, far surpassing the capabilities of existing ground-based and space-based observatories \cite{Labeyrie2009}.

The lunar surface offers several characteristics that are particularly favorable for hypertelescope deployment. The absence of atmospheric turbulence eliminates one of the principal limitations affecting optical interferometry on Earth. In addition, the Moon provides a mechanically stable platform where large arrays of mirrors could be distributed over kilometer-scale distances while maintaining precise relative positioning.

Several conceptual studies have suggested that future lunar hypertelescopes could eventually achieve direct imaging of stellar surfaces, circumstellar environments, accretion disks around compact objects, and potentially even Earth-like exoplanets orbiting nearby stars \cite{Labeyrie2009, Labeyrie2013}.

Although such facilities remain well beyond current engineering capabilities, they illustrate the extraordinary scientific potential that lunar astronomy could offer in the longer term. In this vision, the Moon would not simply host larger telescopes, but entirely new classes of astronomical instruments capable of reaching angular resolutions previously considered unattainable.

\subsection{From Ground Demonstrators to Lunar Interferometers}
\label{sec:7.1}

The concept of using the Moon as a platform for optical interferometry is not entirely new. Over the past decades, several pioneering efforts have sought to demonstrate the feasibility of diluted-aperture imaging and large-baseline optical interferometry on Earth before considering deployment in space or on the lunar surface.

Beyond the successful demonstrations achieved by major Optical Long Baseline Interferometry (OLBI) facilities such as the VLTI and CHARA Array, one of the most notable experimental efforts is the Hypertelescope concept developed by the Association Hypertelescope under the leadership of Antoine Labeyrie. Ground-based demonstrators installed at the Moutière site in the French Alps have successfully validated several key technologies, including pupil densification, coherent beam combination, and high-angular-resolution imaging with sparse optical arrays. These experiments constitute the first practical demonstrations of hypertelescope principles and provide an important technological foundation for future very-large-baseline interferometric observatories, both in space and on the lunar surface.\footnote{\url{https://hypertelescope.org/}}

More recently, renewed international interest in lunar astronomy has led to the emergence of dedicated interferometric concepts specifically designed for deployment on the Moon. One of the most advanced proposals is the \textit{MoonLITE} (Lunar InTerferometry Explorer) mission, submitted within NASA's Astrophysics Pioneers program. MoonLITE proposes the deployment of a two-element optical interferometer on the lunar surface using NASA's Commercial Lunar Payload Services (CLPS) architecture. The concept would employ a lander and rover system to position two telescopes separated by approximately 100~m, enabling sub-milliarcsecond angular resolution and unprecedented sensitivity for optical interferometry from the lunar surface \cite{vanBelle2024MoonLITE}. The mission would constitute the first separated-aperture optical interferometer ever operated beyond Earth \cite{vanBelle2024MoonLITE}.

An even more ambitious concept is the \textit{Artemis-enabled Stellar Imager} (AeSI), currently under study through NASA's Innovative Advanced Concepts (NIAC) program. AeSI envisions a reconfigurable long-baseline ultraviolet and optical interferometer deployed on the lunar surface and supported by the infrastructure developed through the Artemis program. Current studies indicate the feasibility of an interferometric array comprising up to fifteen telescopes distributed over baselines approaching one kilometer, with future expansion toward even larger arrays. Such an observatory would enable direct imaging of stellar surfaces, active galactic nuclei, accretion disks, and potentially surface structures on nearby exoplanets, achieving angular resolutions far beyond those accessible to existing space telescopes \cite{Rau2024AeSI,Carpenter2025AeSI}.

Together, these initiatives illustrate the transition from terrestrial proof-of-concept experiments toward realistic lunar interferometric observatories. They demonstrate that the combination of modern interferometric techniques, robotic deployment capabilities, and emerging lunar infrastructure may finally enable the realization of the long-standing vision of lunar optical interferometry.

\section{FUTURE AMBITIOUS PROJECTS}
\label{sec:8}

Beyond the first generation of lunar observatories currently under study, the Moon could ultimately enable astronomical facilities of unprecedented scale and sensitivity \cite{Dittus2010LunarObservatories}. The unique combination of a stable surface, absence of atmosphere, low seismic activity, and access to permanently shadowed regions offers opportunities that are difficult or impossible to realize either on Earth or in free space \cite{Angel2008LunarTelescope}.

\subsection{Giant Lunar Telescopes for Cryogenic Infrared Astronomy}
\label{sec:8.1}

One particularly promising concept involves the deployment of large-aperture cryogenic infrared telescopes within permanently shadowed craters near the lunar poles. These locations naturally maintain extremely low temperatures, providing an ideal environment for deep infrared observations without the complexity and mass penalties associated with active cryogenic cooling systems \cite{Lester2004LIT}.

The primary scientific objectives of such facilities would include the detection and characterization of the first galaxies formed after cosmic reionization, direct observations of exoplanets and their atmospheres, studies of star and planet formation, and measurements of the Cosmic Infrared Background (CIB) \cite{LUVOIR2019}. Operating at temperatures approaching a few tens of kelvin would dramatically improve sensitivity in the mid- and far-infrared domains.

Several conceptual studies have explored this possibility, including the Lunar Infrared Telescope (LIT) \cite{Lester2004LIT}, various lunar-adapted versions of the Large UV/Optical/Infrared Surveyor (LUVOIR) \cite{LUVOIR2019}, and the FarSide Explorer concept \cite{Burns2013LunarFarside}. Common design characteristics include:

\begin{itemize}
    \item Passive cooling through emplacement within permanently shadowed craters;
    \item Operating temperatures approaching $\sim$30~K or below;
    \item Large primary mirrors ($\geq$4~m), potentially assembled from modular segments;
    \item Long-duration autonomous operations with robotic maintenance capabilities;
    \item Synergies with future lunar infrastructure and power-generation systems.
\end{itemize}

\subsection{Toward Giant Lunar Telescopes}
\label{sec:8.2}

The lunar environment may also revive the long-standing ambition of constructing extremely large optical telescopes exceeding the dimensions currently achievable on Earth. The cancellation of the 100-m class OverWhelmingly Large Telescope (OWL) concept eventually led to the development of ESO's 39-m Extremely Large Telescope (ELT). However, the reduced lunar gravity, absence of atmospheric turbulence, and availability of natural crater topography may permit the construction of much larger apertures in the future \cite{Angel2008LunarTelescope}.

Conceptual studies have proposed lunar telescopes with apertures ranging from 50 to 100~m, potentially exploiting crater geometries, cable-supported structures, modular mirrors, or in-situ manufactured components. Such facilities could bridge the gap between conventional giant telescopes and sparse-aperture interferometric arrays, offering unprecedented sensitivity and angular resolution for studies of exoplanets, stellar populations, and the distant Universe \cite{Dittus2010LunarObservatories}.

While these projects remain highly speculative, ongoing progress in lunar transportation, robotics, autonomous assembly, in-situ resource utilization (ISRU), and advanced power systems suggests that giant lunar observatories may eventually become a realistic component of the long-term astronomical roadmap \cite{Dittus2010LunarObservatories}.

Farside lunar concepts also extend into low-frequency radio astronomy, where shielding from Earth’s radio emissions enables unique observational windows, as discussed in the FarView and FARSIDE studies \cite{Burns2021Farside, Burns2013LunarFarside}.

\section{A PROPOSED ROADMAP TOWARD LUNAR HYPERTELESCOPES}
\label{sec:9}

The realization of kilometer-scale lunar hypertelescopes will likely require a progressive development strategy extending over several decades. Such an approach reduces technical risks while allowing gradual validation of critical technologies under actual lunar conditions.

A realistic roadmap can be divided into four successive phases, each building upon the capabilities demonstrated during the previous stage.

\subsection{Phase I: Robotic Demonstrators}
\label{sec:9.1}

The first phase consists of deploying small robotic astronomical payloads near established lunar landing sites. These precursor missions would primarily serve as technology demonstrators and environmental characterization platforms.

Key objectives include:

\begin{itemize}
\item Characterization of the lunar optical environment;
\item Measurement of dust transport and deposition processes;
\item Evaluation of thermal stability and local seeing conditions;
\item Validation of autonomous operations and robotic deployment procedures;
\item Testing of communication and data-transfer infrastructures.
\end{itemize}

Several ongoing lunar programs, including Artemis and CLPS, may provide opportunities for such pathfinder missions \cite{NASAArtemis2023,CLPS2024}.

\subsection{Phase II: Medium Aperture Observatories}
\label{sec:9.2}

Following successful robotic demonstrations, dedicated scientific observatories with apertures ranging from approximately 1 to 10 meters could be deployed.

These facilities would constitute the first true lunar astronomical observatories capable of conducting competitive scientific programs. Potential instruments include optical, infrared, ultraviolet, and radio telescopes optimized for the unique lunar environment.

Primary objectives would include:

\begin{itemize}
\item Long-duration autonomous scientific operations;
\item Validation of robotic maintenance procedures;
\item Characterization of instrument aging in the lunar environment;
\item Demonstration of precision pointing and tracking systems;
\item Production of significant scientific data sets.
\end{itemize}

Such facilities would provide critical operational experience while establishing the scientific value of routine lunar observations.

\subsection{Phase III: Interferometric Arrays}
\label{sec:9.3}

The third phase would introduce optical and infrared interferometry on the lunar surface.

Multiple telescopes distributed over baselines ranging from tens to hundreds of meters would be connected through coherent beam-combination systems. This stage represents a major technological milestone because it requires nanometric optical path control and high-precision metrology.

Major objectives include:

\begin{itemize}
\item Precision laser metrology;
\item Coherent beam combination;
\item Phase stabilization over large baselines;
\item High-angular-resolution imaging;
\item Validation of distributed observatory architectures.
\end{itemize}

The absence of atmospheric turbulence may provide significant advantages compared with terrestrial interferometric facilities \cite{Lawson2000,Labeyrie2009}.

\subsection{Phase IV: Kilometer-Scale Lunar Hypertelescopes}
\label{sec:9.4}

The final stage would involve the deployment of kilometer-scale diluted optical arrays operating as true hypertelescopes.

Such facilities could consist of hundreds or even thousands of individual collecting elements distributed across large regions of the lunar surface. Advanced pupil densification and beam-combination systems would enable direct imaging at unprecedented angular resolutions.

Scientific objectives could include:

\begin{itemize}
\item Direct imaging of Earth-like exoplanets;
\item Detection of biosignatures in nearby planetary systems;
\item Mapping stellar photospheres and magnetic activity;
\item Imaging environments surrounding black holes;
\item Ultra-high-resolution studies of galaxy evolution and cosmology.
\end{itemize}

Although such instruments remain beyond current engineering capabilities, they represent a logical long-term extension of existing developments in optical interferometry and lunar infrastructure \cite{Labeyrie1996,Labeyrie2013}.

\subsection{Hybrid Lunar Hyper-Telescope Concept}
\label{sec:9.5}

\begin{figure}[htbp]
    \centering
    \includegraphics[width=\textwidth]{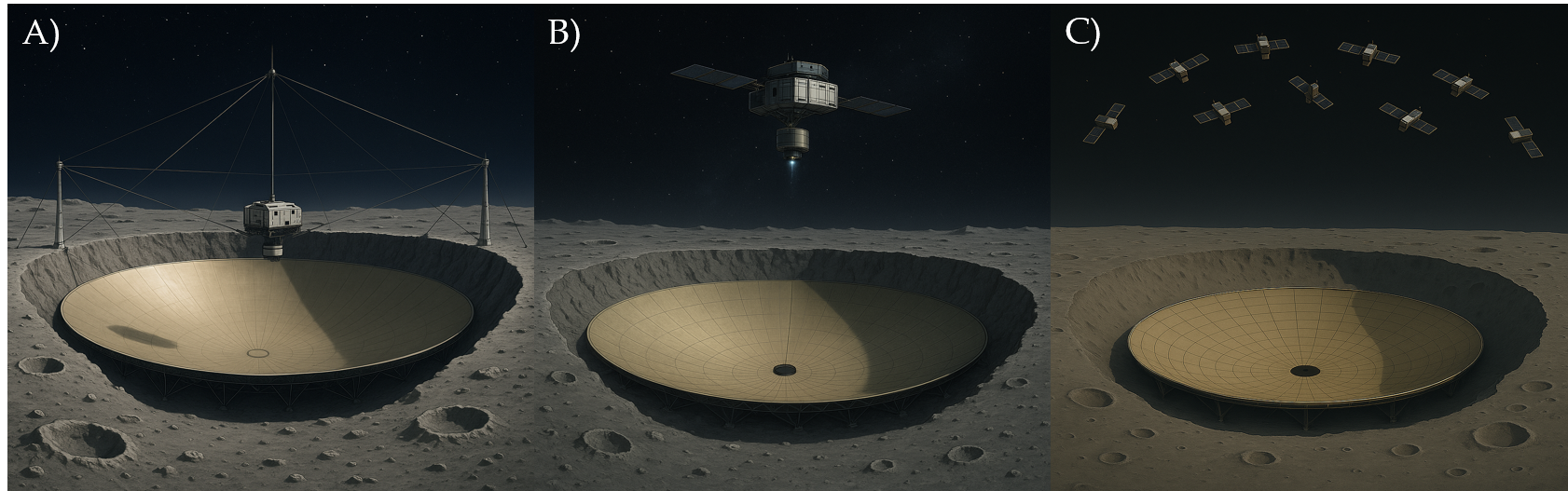}
    \caption{Hybrid lunar hyper-telescope concept: (A) using a suspended cable-controlled nacelle; (B) optical/infrared stabilized platform; (C) interferometric constellation of micro-satellites forming long-baseline observations.}
    \label{fig:1}
\end{figure}

Figure \ref{fig:1} shows the proposed hybrid lunar hyper-telescope architecture based on a variable-geometry, cable-suspended platform deployed above a natural lunar crater. The concept combines proven large-scale radio telescope mechanics (e.g., FAST and Arecibo) with modern robotic, optical, and interferometric technologies, while fully exploiting the unique environmental conditions of the lunar surface.

The system is designed as an evolvable observational infrastructure, enabling a gradual transition from radio astronomy experiments to large-scale optical and infrared observations, and ultimately to kilometer-scale interferometric arrays.

The proposed architecture offers three operational concepts, ranging from the simplest to the most advanced:

\begin{itemize}
\item A suspended robotic nacelle positioned above the crater, operating as a large-aperture optical collector with cable-controlled positioning and alignment.
\item A stabilized orbital or suspended platform enabling high-precision photometric and imaging observations through adaptive focusing, active thermal control, and precision pointing systems.
\item A distributed constellation of micro-satellites or micro-probes forming a long-baseline interferometric array, coherently combining signals collected directly or reflected by the crater-based primary structure.
\end{itemize}

Cables are anchored into the lunar regolith and connected to mast or pillar structures distributed around the crater rim. These masts incorporate controlled tension systems and metrology instruments (laser ranging, GNSS-like local navigation, inertial sensors), ensuring precise geometrical control of the overall structure.

A suspended robotic nacelle (mass range: $\sim$100 kg to 1 ton) provides instrument flexibility, including radio receivers, optical combiners, and active control systems. Optional circular rails installed on the crater rim allow dynamic repositioning of cable anchor points, enabling real-time modulation of the effective aperture geometry.

Segmented reflective surfaces integrated into the crater walls can further enhance optical performance by increasing effective collecting area and enabling configurable optical modes.

Over long-term evolution (50+ years), the system can naturally evolve into a lunar interferometric observatory, where multiple coordinated elements operate coherently with baselines ranging from tens to hundreds of kilometers.

\section{SCIENTIFIC OPPORTUNITIES}
\label{sec:10}

Lunar hypertelescopes could enable scientific investigations that are currently inaccessible to both ground-based observatories and existing space telescopes.

Their unprecedented angular resolution would open entirely new observational domains across multiple fields of astrophysics.

\subsection{Exoplanet Imaging}
\label{sec:10.1}

One of the most compelling scientific motivations is the direct imaging of Earth-like planets orbiting nearby stars. Kilometer-scale baselines could potentially resolve planetary systems at angular scales sufficient to separate habitable-zone planets from their parent stars and perform spectroscopic analyses of their atmospheres \cite{Labeyrie1996,Labeyrie2009}.

\subsection{Stellar Astrophysics}
\label{sec:10.2}

Lunar hypertelescopes could resolve the surfaces of large numbers of stars beyond the Sun. Detailed imaging of stellar photospheres, convection cells, starspots, magnetic structures, and mass-loss processes would become possible, providing unprecedented tests of stellar evolution models.

\subsection{Black Hole Physics}
\label{sec:10.3}

The environments surrounding supermassive black holes represent another major scientific target. Future lunar interferometric facilities may complement projects such as the Event Horizon Telescope by providing observations at shorter wavelengths and significantly higher angular resolutions \cite{EHT2019}.

\subsection{Cosmology and Galaxy Evolution}
\label{sec:10.4}

Ultra-high-resolution observations of distant galaxies could reveal the internal structure of early galactic systems, active galactic nuclei, and gravitational lens configurations. Such observations would provide new constraints on galaxy formation and cosmological evolution.

\subsection{Solar System Science}
\label{sec:10.5}

The Solar System would also benefit enormously from lunar hypertelescope capabilities. Detailed surface mapping of asteroids, comets, trans-Neptunian objects, and planetary satellites could become routine, providing valuable information about Solar System formation and evolution.

Collectively, these scientific opportunities illustrate how lunar hypertelescopes could transform astronomy in much the same way that large optical telescopes revolutionized the field during the twentieth century.

\section{LONG-TERM VISION: A LUNAR ASTRONOMY NETWORK}
\label{sec:11}

In the longer term, lunar astronomy may evolve beyond individual facilities toward an integrated network of observatories distributed across the lunar surface.

Such a network would exploit the diversity of lunar environments and support a broad range of scientific objectives while sharing common infrastructure, communications, power systems, and maintenance resources.

Potential components could include:

\begin{itemize}
\item Optical observatories located on elevated terrain;
\item Infrared observatories operating within permanently shadowed regions;
\item Low-frequency radio observatories on the lunar far side;
\item Optical and infrared interferometric arrays;
\item Kilometer-scale hypertelescope infrastructures.
\end{itemize}

The lunar far side is particularly attractive for radio astronomy due to its natural shielding from terrestrial radio-frequency interference, while permanently shadowed craters may provide exceptionally cold environments favorable for infrared observations \cite{Burns2013FARSIDE,Lester2020}.

As lunar exploration infrastructure matures, individual observatories could become interconnected through high-bandwidth communication systems and coordinated observation strategies. Such an architecture would resemble a planetary-scale observatory capable of addressing scientific questions across the entire electromagnetic spectrum.

Rather than replacing terrestrial or free-space observatories, a lunar astronomy network would operate as a complementary component of a broader astronomical ecosystem. Ground-based facilities, Earth-orbiting telescopes, deep-space observatories, and lunar installations would each contribute unique capabilities optimized for specific scientific objectives.

Ultimately, the establishment of a permanent lunar astronomy network could represent one of the most significant expansions of humanity's observational capabilities since the advent of modern astronomy. The Moon would evolve from an object of astronomical study into a major platform for astronomical discovery.

\section{CONCLUSIONS}
\label{sec:Conclu}

The concept of lunar astronomy has undergone a remarkable evolution over the past six decades. What began during the Apollo era as a collection of visionary and often highly speculative ideas is progressively emerging as a realistic long-term objective supported by major international lunar programs, advances in robotics, and the rapid development of astronomical instrumentation.

Historically, numerous studies identified the Moon as an exceptional location for astronomical observations. The absence of a significant atmosphere, the availability of stable geological platforms, the radio-quiet environment of the lunar far side, and the potential for constructing infrastructures on scales unattainable in Earth orbit all motivated a wide range of proposals. However, despite the scientific appeal of these concepts, early lunar observatories never progressed beyond the conceptual stage. Technical limitations, high transportation costs, the absence of a permanent lunar infrastructure, and shifting political priorities following Apollo ultimately prevented their realization.

Today, the situation is fundamentally different. The renewed international interest in lunar exploration, driven by programs such as Artemis, the International Lunar Research Station (ILRS), and a growing commercial lunar sector, is creating the foundations for a sustained human and robotic presence on the Moon. At the same time, major advances in autonomous systems, precision metrology, additive manufacturing, in-situ resource utilization, lightweight optics, and optical interferometry are steadily removing many of the barriers that constrained earlier generations of lunar astronomy concepts.

These developments are opening the door to a new class of astronomical facilities. Initial robotic demonstrators and medium-aperture observatories may eventually evolve into distributed interferometric arrays operating over baselines of hundreds of meters or even several kilometers. Such facilities would exploit the unique characteristics of the lunar environment while complementing both terrestrial observatories and future space-based missions.

Among the most ambitious concepts considered to date, lunar hypertelescopes stand out as a particularly compelling long-term objective. By combining light collected from large numbers of distributed apertures through coherent beam combination and pupil densification techniques, these instruments could achieve angular resolutions far beyond those of conventional telescopes. Their scientific potential includes direct imaging of Earth-like exoplanets, detailed mapping of stellar surfaces, studies of black-hole environments, investigations of galaxy evolution, and a broad range of transformative observations that remain inaccessible to current facilities.

The development of such observatories will undoubtedly require sustained international cooperation, long-term investment, and decades of technological maturation. Nevertheless, the progression from robotic pathfinders to large-scale interferometric infrastructures follows a plausible and incremental roadmap consistent with current trends in both lunar exploration and astronomical instrumentation.

Ultimately, the most profound implication of lunar astronomy may not be the construction of larger telescopes, but the emergence of an entirely new observational paradigm. For centuries, the Moon has served as one of humanity's most important astronomical targets. During the twenty-first century, it may gradually become one of humanity's most powerful astronomical platforms. The transformation of the Moon from an object of observation into a foundation for observation could represent one of the defining scientific achievements of the coming space age.

\bibliography{report} 

@book{galilei1610sidereus,
  title     = {Sidereus Nuncius},
  author    = {Galilei, Galileo},
  year      = {1610},
  publisher = {Thomas Baglioni},
  address   = {Venice}
}

@book{north2008cosmos,
  title     = {Cosmos: An Illustrated History of Astronomy and Cosmology},
  author    = {North, John},
  year      = {2008},
  publisher = {University of Chicago Press},
  address   = {Chicago}
}

@book{newton1687principia,
  author    = {Newton, Isaac},
  title     = {Philosophiae Naturalis Principia Mathematica},
  year      = {1687},
  publisher = {Royal Society},
  address   = {London}
}

@book{compton1989apollo,
  author    = {Compton, William David},
  title     = {Where No Man Has Gone Before: A History of {Apollo} Lunar Exploration Missions},
  series    = {{NASA} History Series},
  number    = {{NASA SP-4214}},
  publisher = {National Aeronautics and Space Administration, Scientific and Technical Information Division},
  year      = {1989},
  address   = {Washington, D.C.}
}

@article{lowman1966lunarresources,
  author  = {Lowman, Jr., Paul D.},
  title   = {Lunar Resources: Their Value in Lunar and Planetary Exploration},
  journal = {Astronautica Acta},
  volume  = {12},
  number  = {5--6},
  pages   = {384--393},
  year    = {1966}
}

@article{Foing1994,
  author   = {Foing, Bernard H.},
  title    = {Astronomy and space science from the moon: Panel discussion and perspectives},
  journal  = {Advances in Space Research},
  year     = {1994},
  volume   = {14},
  number   = {6},
  pages    = {283--288},
  doi      = {10.1016/0273-1177(94)90039-6},
  adsurl   = {https://ui.adsabs.harvard.edu/abs/1994AdSpR..14f.283F}
}

@proceedings{Foing1994Proc,
  editor       = {Foing, Bernard H.},
  title        = {Astronomy and Space Science from the Moon},
  series       = {Advances in Space Research},
  volume       = {14},
  number       = {6},
  year         = {1994},
  organization = {COSPAR},
  publisher    = {Pergamon Press},
  note         = {Proceedings of Symposium E4 of the COSPAR 29th Plenary Meeting, Washington, DC, USA, 28 August--5 September 1992},
  adsurl       = {https://ui.adsabs.harvard.edu/abs/1994AdSpR..14f....F}
}

@misc{NASAArtemis2023,
  author       = {{NASA}},
  title        = {{NASA} {Artemis} Program Overview},
  year         = {2023},
  howpublished = {\url{https://www.nasa.gov/artemis}},
  note         = {Accessed: 2026-07-05}
}

@article{Wu2021ILRS,
  author  = {Wu, Weiren and Wang, Qiong and Tang, Yuhua and Zhang, Rongqiao and Zhang, Tianyi},
  title   = {International Lunar Research Station: A New Model for Lunar Exploration},
  journal = {Chinese Journal of Space Science},
  year    = {2021},
  volume  = {41},
  number  = {5},
  pages   = {635--645}
}

@misc{CNSA2021,
  author       = {{China National Space Administration}},
  title        = {Guide to the International Lunar Research Station},
  year         = {2021},
  howpublished = {Official Report}
}

@misc{CLPS2024,
  author       = {{NASA}},
  title        = {Commercial Lunar Payload Services ({CLPS})},
  year         = {2024},
  howpublished = {\url{https://www.nasa.gov/clps}},
  note         = {Accessed: 2026-07-05}
}

@inproceedings{Fong2008,
  author    = {Fong, Terrence and Bualat, Maria and Deans, Matt and Adams, Byron and To, Vinh},
  title     = {Field Testing of Human-Robot Systems for Lunar Exploration},
  booktitle = {AIAA SPACE 2008 Conference \& Exposition},
  year      = {2008},
  pages     = {7753},
  publisher = {AIAA}
}

@inproceedings{Biesiadecki2020,
  author    = {Biesiadecki, Jeffrey and Rothrock, Brandon and Maimone, Mark and San Martin, Alejandro},
  title     = {Mars 2020 Rover Autonomy and Robotic Technologies},
  booktitle = {2020 IEEE Aerospace Conference},
  year      = {2020},
  pages     = {1--12},
  publisher = {IEEE}
}

@article{Anand2012,
  author  = {Anand, M. and Crawford, I. A. and Balat-Pichelin, M. and Abanades, S. and Westrenen, W. van and Pernet-Fisher, J. and Duchêne, I. and Flahaut, J. and Olsson-Francis, K.},
  title   = {A Brief Review of Chemical and Mineralogical Resources on the Moon and Their Potential Utilization},
  journal = {Planetary and Space Science},
  volume  = {74},
  number  = {1},
  pages   = {42--48},
  year    = {2012}
}

@techreport{SandersLarson2015,
  author      = {Sanders, Gerald B. and Larson, William E.},
  title       = {Progress Made in Lunar In Situ Resource Utilization Under {NASA}'s Exploration Technology and Development Program},
  institution = {National Aeronautics and Space Administration},
  year        = {2015},
  number      = {NASA/TM-2015-218844}
}

@article{Wang2015LUT,
  author  = {Wang, Jian-Ning and Deng, Licai and Cao, Lingzhi and Han, Yanbiao and Wei, Jianyan and Cao, Li and Urata, Yuji},
  title   = {Photometric Performance of the Lunar-based Ultraviolet Telescope on {Chang'e-3}},
  journal = {Research in Astronomy and Astrophysics},
  volume  = {15},
  number  = {7},
  pages   = {1068--1080},
  year    = {2015}
}

@article{Cao2017LUT,
  author  = {Cao, Lingzhi and Deng, Licai and Wang, Jianning and Han, Yanbiao and Meng, Xiaoming and Wei, Jianyan},
  title   = {The First Lunar-based Ultraviolet Telescope: Scientific Results and Performance},
  journal = {Publications of the Astronomical Society of the Pacific},
  volume  = {129},
  number  = {972},
  pages   = {024401},
  year    = {2017}
}

@techreport{Burns2021LCRT,
  author      = {Burns, Jack O. and Hallinan, Gregg and Bienfang, Joshua and Bowman, Judd and Bradley, Richard and Mirocha, Jordan},
  title       = {Lunar Crater Radio Telescope ({LCRT}): {NIAC} Phase {II} Final Report},
  institution = {NASA Innovative Advanced Concepts},
  year        = {2021},
  type        = {Final Report}
}

@article{Burns2020DAPPER,
  author  = {Burns, Jack O. and Tauscher, Keith and Rapetti, David and Bowman, Judd D. and Bradley, Richard and Mirocha, Jordan and Furlanetto, Steven R.},
  title   = {A Space-Based Observational Strategy for Characterizing the First Stars and Galaxies Using the Redshifted 21-cm Global Spectrum},
  journal = {The Astrophysical Journal},
  volume  = {844},
  number  = {1},
  pages   = {33},
  year    = {2020}
}

@article{Lester2020,
  author  = {Lester, Dan and Thronson, Harley and MacEwen, Howard},
  title   = {Scientific Opportunities for Infrared Astronomy from the Lunar Poles},
  journal = {Planetary and Space Science},
  volume  = {180},
  pages   = {104750},
  year    = {2020}
}

@article{Angel2008,
  author  = {Angel, Roger and Worden, Simon and Borra, Ermanno F. and Eisenstein, Daniel and Finkelstein, Steven and Green, Richard F. and Hubeny, Ivan and Ma, Zhiwei and McCarthy, Don and Meyer, Michael and others},
  title   = {Lunar Liquid Mirror Telescopes and the Origins of the Universe},
  journal = {Astrophysics and Space Science},
  volume  = {320},
  number  = {1},
  pages   = {197--203},
  year    = {2008}
}

@article{Cesaretti2014,
  author  = {Cesaretti, Giorgio and Dini, Enrico and De Kestelier, Xavier and Colla, Valentina and Pambaguian, Laurent},
  title   = {Building Components for an Outpost on the Lunar Soil by Means of a Novel {3D} Printing Technology},
  journal = {Acta Astronautica},
  volume  = {93},
  pages   = {430--450},
  year    = {2014}
}

@article{Meurisse2018,
  author  = {Meurisse, Alexandre and Makaya, Advenit and Willsch, Christoph and Sperl, Martin},
  title   = {Solar {3D} Printing of Lunar Regolith},
  journal = {Acta Astronautica},
  volume  = {152},
  pages   = {800--810},
  year    = {2018}
}

@article{Angel2003,
  author  = {Angel, J. Roger P.},
  title   = {Large Optical Telescopes and the Future of Astronomy},
  journal = {Nature},
  volume  = {423},
  number  = {6942},
  pages   = {915--921},
  year    = {2003}
}

@book{Lawson2000,
  author    = {Lawson, Peter R.},
  title     = {Principles of Long Baseline Stellar Interferometry},
  publisher = {NASA/JPL, California Institute of Technology},
  year      = {2000},
  address   = {Pasadena, CA}
}

@article{BlandHawthorn2019,
  author  = {Bland-Hawthorn, Joss and Kern, Andreas},
  title   = {Astrophotonics: A New Era for Astronomical Instruments},
  journal = {Optica},
  volume  = {6},
  number  = {5},
  pages   = {646--656},
  year    = {2019}
}

@misc{WNN_thorium_2025,
  author       = {{World Nuclear News}},
  title        = {Chinese molten salt reactor achieves conversion of thorium-uranium fuel},
  year         = {2025},
  howpublished = {\url{https://www.world-nuclear-news.org/articles/chinese-msr-achieves-conversion-of-thorium-uranium-fuel}},
  note         = {Accessed: 2026-07-05}
}

@misc{WNN_operational_2023,
  author       = {{World Nuclear News}},
  title        = {Operating permit issued for Chinese molten salt reactor},
  year         = {2023},
  howpublished = {\url{https://www.world-nuclear-news.org/Articles/Operating-permit-issued-for-Chinese-molten-salt-re}},
  note         = {Accessed: 2026-07-05}
}

@misc{CAS_thorium_2025,
  author       = {{Chinese Academy of Sciences (SINAP)}},
  title        = {China achieves thorium-uranium nuclear fuel conversion in molten salt reactor},
  year         = {2025},
  howpublished = {\url{https://english.cas.cn/newsroom/cas_media/202511/t20251104_1095922.shtml}},
  note         = {Accessed: 2026-07-05}
}

@misc{Moon_Nuclear_ILRS_2025,
  author       = {{The Times}},
  title        = {China plans nuclear power plant on the Moon for {ILRS} program},
  year         = {2025},
  howpublished = {\url{https://www.thetimes.com/world/asia/article/china-moon-nuclear-power-plant-russia-rprwclfb2}},
  note         = {Accessed: 2026-07-05}
}

@article{Gardner2006,
  author  = {Gardner, Jonathan P. and Mather, John C. and Clampin, Mark and Doyon, René and Greenhouse, Matthew A. and Hammel, Heidi B. and Hutchings, John B. and Jakobsen, Peter and Lilly, Simon J. and Long, Knox S. and others},
  title   = {The {James Webb Space Telescope}},
  journal = {Space Science Reviews},
  volume  = {123},
  number  = {4},
  pages   = {485--606},
  year    = {2006}
}

@article{Mather2022,
  author  = {Mather, John C. and others},
  title   = {The {James Webb Space Telescope} Mission},
  journal = {Nature Astronomy},
  volume  = {6},
  number  = {1},
  pages   = {121--129},
  year    = {2022}
}

@article{Burns2013FARSIDE,
  author  = {Burns, Jack O. and Lazio, Joseph and Bale, Stuart and Bowman, Judd and Bradley, Richard and Carilli, Chris and Furlanetto, Steven and Harker, Geraint and Loeb, Abraham and Mirocha, Jordan and others},
  title   = {Probing the First Stars and Black Holes in the Early Universe with the {Dark Ages Radio Explorer} and the Lunar Farside},
  journal = {Advances in Space Research},
  volume  = {52},
  number  = {2},
  pages   = {306--320},
  year    = {2013}
}

@article{Jester2009,
  author  = {Jester, Sebastian and Falcke, Heino},
  title   = {Science with a Lunar Low-Frequency Array: From the Dark Ages of the Universe to Nearby Exoplanets},
  journal = {New Astronomy Reviews},
  volume  = {53},
  number  = {1--4},
  pages   = {1--26},
  year    = {2009}
}

@article{Borra2007,
  author  = {Borra, Ermanno F.},
  title   = {Liquid Mirror Telescopes on the Moon},
  journal = {Advances in Space Research},
  volume  = {40},
  number  = {1},
  pages   = {125--133},
  year    = {2007}
}

@inproceedings{Zacny2019,
  author    = {Zacny, Kris and Mueller, Robert and King, Jack and Hedlund, Michael and Craft, John},
  title     = {Technologies for Large Lunar Surface Infrastructure and Observatory Construction},
  booktitle = {AIAA SPACE and Astronautics Forum and Exposition},
  year      = {2019},
  pages     = {4091},
  publisher = {AIAA}
}

@article{Labeyrie1996,
  author  = {Labeyrie, Antoine},
  title   = {Resolved Imaging of Extra-Solar Planets with Future 10--100 km Optical Interferometric Arrays},
  journal = {Astronomy and Astrophysics Supplement Series},
  volume  = {118},
  number  = {3},
  pages   = {517--524},
  year    = {1996}
}

@book{Labeyrie2013,
  author    = {Labeyrie, Antoine and Lipson, Stephen G. and Nisenson, Peter},
  title     = {Hypertelescopes},
  publisher = {Springer},
  year      = {2013},
  address   = {Dordrecht}
}

@article{Labeyrie2009,
  author  = {Labeyrie, Antoine},
  title   = {Towards Extremely Large Diluted Aperture Telescopes},
  journal = {Experimental Astronomy},
  volume  = {23},
  number  = {2},
  pages   = {571--603},
  year    = {2009}
}

@article{EHT2019,
  author  = {{Event Horizon Telescope Collaboration}},
  title   = {First {M87} {Event Horizon Telescope} Results. {I}. The Shadow of the Supermassive Black Hole},
  journal = {The Astrophysical Journal Letters},
  volume  = {875},
  number  = {1},
  pages   = {L1},
  year    = {2019}
}

@article{vanBelle2024MoonLITE,
  author = {van Belle, Gerard T. and Ciardi, David and Hillsberry, Daniel and others},
  title = {MoonLITE: a CLPS-delivered NASA Astrophysics Pioneers lunar optical interferometer for sensitive, milliarcsecond observing},
  journal = {Proceedings of SPIE},
  volume = {13092},
  pages = {130922N},
  year = {2024},
  eprint = {2408.01392},
  archivePrefix = {arXiv},
  primaryClass = {astro-ph.IM}
}

@article{Rau2024AeSI,
  author = {Rau, Gioia and Carpenter, Kenneth G. and Boyajian, Tabetha and others},
  title = {Artemis-enabled Stellar Imager (AeSI): A Lunar Long-Baseline UV/Optical Imaging Interferometer},
  journal = {Proceedings of SPIE},
  year = {2024},
  eprint = {2408.04699},
  archivePrefix = {arXiv},
  primaryClass = {astro-ph.IM}
}

@article{Carpenter2025AeSI,
  author = {Carpenter, Kenneth G. and Boyajian, Tabetha and Buzasi, Derek and others},
  title = {NASA Innovative Advanced Concepts Phase I Final Report -- A Lunar Long-Baseline UV/Optical Imaging Interferometer: Artemis-enabled Stellar Imager (AeSI)},
  journal = {arXiv preprint arXiv:2503.02105},
  year = {2025},
  eprint = {2503.02105},
  archivePrefix = {arXiv},
  primaryClass = {astro-ph.IM}
}

@article{Burns2013LunarFarside,
  author = {Burns, Jack O. and Lazio, Joseph and Bale, Stuart and others},
  title = {A Lunar L2-Farside Exploration and Science Mission Concept with the Orion Multi-Purpose Crew Vehicle and a Teleoperated Lander/Rover},
  journal = {Advances in Space Research},
  volume = {52},
  number = {2},
  pages = {306--320},
  year = {2013},
  doi = {10.1016/j.asr.2012.06.032}
}

@article{Lester2004LIT,
  author = {Lester, Daniel F. and Smith, Howard A. and others},
  title = {The Lunar Infrared Telescope (LIT): A Concept for Infrared Astronomy from the Moon},
  journal = {Advances in Space Research},
  volume = {34},
  number = {4},
  pages = {568--574},
  year = {2004},
  doi = {10.1016/j.asr.2003.02.059}
}

@article{LUVOIR2019,
  author = {{The LUVOIR Team}},
  title = {The LUVOIR Mission Concept Study Final Report},
  journal = {arXiv preprint arXiv:1912.06219},
  year = {2019},
  archivePrefix = {arXiv},
  eprint = {1912.06219},
  primaryClass = {astro-ph.IM}
}

@article{Angel2008LunarTelescope,
  author = {Angel, J. Roger P. and Woolf, Neville J.},
  title = {Astronomical Telescopes on the Moon},
  journal = {Science},
  volume = {320},
  number = {5881},
  pages = {1503--1504},
  year = {2008},
  doi = {10.1126/science.1158241}
}

@article{Burns2021Farside,
  author = {Burns, Jack O. and others},
  title = {Farside Radio Astronomy on the Moon: The FarView and FARSIDE Concepts},
  journal = {Planetary and Space Science},
  volume = {204},
  pages = {105268},
  year = {2021},
  doi = {10.1016/j.pss.2021.105268}
}

@article{Dittus2010LunarObservatories,
  author = {Dittus, Hans and others},
  title = {Astronomy from the Moon: The Next Decades},
  journal = {Experimental Astronomy},
  volume = {26},
  number = {1-3},
  pages = {151--197},
  year = {2010},
  doi = {10.1007/s10686-009-9158-5}
}
\bibliographystyle{spiebib} 

\end{document}